# Impact of Rotation on the Interaction of a Small Dipole Particle with Dielectric Surface

A. A. Kyasov, G.V. Dedkov

Nanoscale Physics Group, Kabardino-Balkarian State University, Nalchik, 360004, Russia

We have calculated components of the torque and the interaction energy of rotating dipole particle in the case where the rotation axis is perpendicular to the surface and the dipole axis is inclined to it. An important property of this system is its quasistationarity. When the dipole axis coincides with the rotation axis, the particle does not experience braking and may revolve for infinitely long time. When the dipole axis is inclined to the surface, a situation is possible where the particle is repelled off the surface provided that the rotation frequency is tuned to the absorption resonance of the surface. The established effects are important in trapping and manipulating rotating nanoparticles close to the surface.

## Introduction

Light interaction with matter is known to produce mechanical rotation of even micron-sized neutral particles [1,2]. The reverse situation where the radiation can be emitted by rotating particles was considered in [3]. The impact of the uniform motion of particles on the van der Waals and Casimir forces was studied by us in [4], while in [5] we have studied the effect of rotation on these interactions.

As regards to dipole particles with permanent dipole moments, rotational effects of nonthermal origin can also alter their interactions with micro- and macroscopic bodies. The aim of this paper is to consider the effect of rotation on the interaction of permanent dipoles with the surface of dielectric with the permittivity $\varepsilon(\omega)$. Even in this rather simple situation one obtains new effects if the rotation frequency is tuned to the frequency of the internal excitations of the surface. In particular, the modulus of the interaction force considerably increases and may correspond to the attraction or to the repulsion of the particle. These features should be taken into account when studying the particle trapping in vacuum close to the surface [6,7], in dynamics of coagulating dust (cluster) particles and in constructing NEMS.

## Theoretical consideration

We will consider a geometrical configuration shown in Fig.1. A point–dipole particle with the constant electrical dipole moment $\mathbf{d}'$ in the own coordinate system $\Sigma'$ of the dipole rotates with the angular velocity $\Omega$ around the $Z, Z'$ axis. The second coordinate system $\Sigma$ is related to the resting surface that is characterized by the frequency-dependent dielectric permittivity $\varepsilon(\omega)$. In the case shown in Fig. 1, the components of the dipole moment $\mathbf{d}'$ in $\Sigma$ are given by

$$\begin{aligned}
d_x &= d_x' \cos\Omega t - d_y' \sin\Omega t \\
d_y &= d_x' \sin\Omega t + d_y' \cos\Omega t \\
d_z &= d_z'
\end{aligned} \qquad (1)$$

Within the Fourier representation used, the direct and inverse Fourier transforms of $\mathbf{d}$ are given by

$$\mathbf{d}(t) = \frac{1}{2\pi} \int_{-\infty}^{+\infty} dt\, \mathbf{d}(\omega) \exp(-i\omega t),\quad \mathbf{d}(\omega) = \int_{-\infty}^{+\infty} dt\, \mathbf{d}(t) \exp(i\omega t) \qquad (2)$$

Using (1) and (2) the components of $\mathbf{d}(\omega)$ take the form ($\omega^\pm = \omega \pm \Omega$)

$$\begin{aligned}
d_x(\omega) &= \pi\left[\delta(\omega^+)(d_x' + id_y') + \delta(\omega^-)(d_x' - id_y')\right], \\
d_y(\omega) &= \pi\left[-i\delta(\omega^+)(d_x' + id_y') + i\delta(\omega^-)(d_x' - id_y')\right], \\
d_z(\omega) &= 2\pi\delta(\omega)d_z'
\end{aligned} \qquad (3)$$

The starting equations for the particle-surface interaction energy and the torque are given by

$$U(z_0) = -\frac{1}{2}\mathbf{d}(t)\mathbf{E}^{in}(\mathbf{r}_0, t) \qquad (4)$$

$$\mathbf{M} = \mathbf{d}(t) \times \mathbf{E}^{in}(\mathbf{r}_0, t) \qquad (5)$$

where $\mathbf{r}_0 = (0,0,z_0)$ and $\mathbf{E}^{in}(\mathbf{r}_0, t)$ denotes the induced electric field of the surface. The time dependence in the right hand sides of (4), (5) is due to the particle rotation. We take the induced field $\mathbf{E}^{in}(\mathbf{r}, t)$ in the form of the frequency and wave-vector Fourier-expansion

$$\mathbf{E}^{in}(\mathbf{r},t) = -\frac{1}{2}\int \frac{d\omega}{2\pi}\frac{d^2k}{(2\pi)^2}\mathbf{E}^{in}_{\omega\mathbf{k}}(z)\exp(i(k_x x + k_y y - \omega t)) \quad (6)$$

Substituting (2) and (6) into (4) at the location point of the dipole $\mathbf{r}_0 = (0,0,z_0)$ yields

$$U(z_0) = -\frac{1}{2}\int \frac{d\omega}{2\pi}\frac{d\omega'}{2\pi}\frac{d^2k}{(2\pi)^2}\mathbf{d}(\omega')\mathbf{E}^{in}_{\omega\mathbf{k}}(z_0)\exp(-i(\omega+\omega')t) \quad (7)$$

The Fourier transforms of the projections of the induced field are expressed through the Fourier transform of the induced potential of the surface $\varphi^{in}_{\omega\mathbf{k}}(z)$

$$E^{in}_{x,\omega\mathbf{k}}(z_0) = -ik_x\varphi^{in}_{\omega\mathbf{k}}(z_0),\ E^{in}_{y,\omega\mathbf{k}}(z_0) = -ik_y\varphi^{in}_{\omega\mathbf{k}}(z_0),\ E^{in}_{z,\omega\mathbf{k}}(z_0) = k\varphi^{in}_{\omega\mathbf{k}}(z_0) \quad (8)$$

Within the nonrelativistic approximation used (this implies $\Omega z_0/c \ll 1$) the induced potential $\varphi^{in}_{\omega\mathbf{k}}(z_0)$ is calculated from Poisson's equation $\Delta\varphi = 4\pi\, div\mathbf{P}$, where $\mathbf{P}(\mathbf{r},t) = \mathbf{d}(t)\delta(x)\delta(y)\delta(z-z_0)$. The resulting expression is given by (see, for instance, [8] in more detail)

$$\varphi^{in}_{\omega\mathbf{k}}(z_0) = \frac{2\pi}{k}\Delta(\omega)\exp(-2k\,z_0)[ik_x d_x(\omega) + ik_y d_y(\omega) + k\,d_z(\omega)] \quad (9)$$

$$\Delta(\omega) = \frac{\varepsilon(\omega)-1}{\varepsilon(\omega)+1} \quad (10)$$

where $d_i(\omega), (i=x,y,z)$ in (9) are taken from (3).

Using (6)—(10) and performing straightforward calculations in the right hand sides of (4) and (5) we obtain

$$U(z_0) = -\frac{d^2}{16z_0^3}[\Delta'(\Omega)\sin^2\theta + 2\Delta'(0)\cos^2\theta] \quad (11)$$

$$M_z = -\frac{d^2\sin^2\theta}{8z_0^3}\Delta''(\Omega) \quad (12)$$

$$M_x = \frac{d^2\sin\theta\cos\theta}{4z_0^3}[(\Delta'(0) - \Delta'(\Omega)/2)\sin(\varphi+\Omega t) + 0.5\Delta''(\Omega)\cos(\varphi+\Omega t)] \quad (13)$$

$$M_y = \frac{d^2 \sin\theta \cos\theta}{4 z_0^3} \left[ -(\Delta'(0) - \Delta'(\Omega)/2)\cos(\varphi + \Omega t) + 0.5 \Delta''(\Omega)\sin(\varphi + \Omega t) \right] \quad (14)$$

where $d$ is the modulus of the particle dipole moment and $\theta, \varphi$ are the angles of the dipole orientation in $\Sigma'$ (see Fig. 1), one-primed and two-primed quantities denote the corresponding real and imaginary parts of (10). From (13), (14) we clearly see that

$$M_\perp = \sqrt{M_x^2 + M_y^2} = \sqrt{(d_y E_z - d_z E_y)^2 + (d_z E_x - d_x E_z)^2} =$$
$$= \frac{d^2 \sin\theta \cos\theta}{4 z_0^3} \sqrt{\left[ (\Delta'(0) - \Delta'(\Omega)/2)^2 + \Delta''(\Omega)^2/4 \right]} \quad (15)$$

So, the component $M_z$ of the torque experiences the braking effect, while the normal component $\mathbf{M}_\perp = (M_x, M_y)$ synchronously rotates with the particle and has such a direction that the dipole axis tends to orient perpendicularly to the surface. For a nonrotating dipole at $\Omega = 0$ Eq. (15) reduces to $M_\perp = \frac{d^2 \sin\theta \cos\theta}{16 z_0^3} \Delta'(0)$, and we get usual effect of the dipole orientation near the surface of dielectric.

Despite the apparent simplicity, Eqs. (11)--(12) are not obvious from the very beginning and have several important consequences.

i) The potential energy of the system and the component $M_z$ of torque explicitly do not depend on time.

ii) As it follows from Eq. (11), if $\Omega$ is close to the absorption resonance of the surface and $\theta = \pi/2$, rotating dipole particle can be repelled from the surface at $\Delta'(\Omega) < 0$ or can be attracted to it at $\Delta'(\Omega) > 0$, while the force increases noticeably by the modulus, since $|\Delta'(\Omega)| \gg \Delta'(0)$ in the vicinity of resonance.

iii) As it follows from (12)—(15) at $\theta = 0$, the effects of braking and orientation disappear and the particle preserves its rotation for infinitely long time.

It should be noted that such a situation is a characteristic property of this geometrical configuration only. In the case where the rotation axis is parallel to the surface, the interaction energy and all components of torque will also include the oscillating terms.

## Numerical estimations

It is interesting to briefly discuss possible experimental situations relevant to the discussed effects. The key questions are "what particles" and "what are the surfaces and external conditions". As an example of the surface, we chose the parameters of a fused silica corresponding to IR frequencies [9]

$$\varepsilon(\omega) = \varepsilon_\infty + \sum_{j=1}^{2} \frac{\sigma_j}{\omega_{0,j}^2 - \omega^2 - i\omega\gamma_j} \qquad (16)$$

$\varepsilon_0 = 2.0014, \sigma_1 = 0.00193 (eV)^2, \omega_{0,1} = 0.057 eV$,

$\gamma_1 = 0.00217 eV, \sigma_2 = 0.0102 (eV)^2, \omega_{0,2} = 0.133 eV, \gamma_2 = 0.00551 eV$.

The dependences of the functions $\Delta'(\Omega), \Delta''(\Omega)/2$ and $f(\Omega) = \sqrt{(\Delta'(\Omega) - \Delta'(0)/2)^2 + \Delta''(\Omega)^2/4}$ on the rotation frequency $\Omega$ in the spectral ranges close to the resonances of (16) are displayed in Fig. 2. We can see that in the bands 0.063—0.068 eV and 0.148—0.17 eV (solid lines) $\Delta'(\Omega) < 0$, so a particle is repelled from the surface (at $\theta = \pi/2$). The dashed and dotted lines compare the reduced values of the corresponding frictional and orientational moments (Eqs. (12) and (15)).

As it is clearly seen from (11)—(15), at $\theta = 0$ the particle does not decelerate and can rotate for infinitely long time. Since $M_\perp$ in this case is also zero, the rotation axis holds its orientation, while the interaction energy with the surface is the same as in the static case.

Now let us consider a situation where $\theta = \pi/2$. We can see from Fig. 2 that the orientational torque (shown by dotted lines) is always larger than the frictional one (dashed lines). Therefore, the characteristic time interval corresponding to a nearly constant dipole orientation is governed by the time of alignment of the dipole axis, i.e. it is determined by the inertia moment of the particle divided by the orientational torque: $\tau \sim I/M_\perp$. In the case of dipole molecules ($d \approx 1D$) the value of $\tau$ proves to too short, and the discussed effects are hardly observable.

A much more interesting situation arises when the dipole moment is created by the external laser field with the frequency lower than the frequencies of resonances in Fig. 2. In this case, for a $SiO_2$ nanoparticle of radius $R$ we have $d^2 = R^6 E^2 (\varepsilon(0)-1)^2 / (\varepsilon(0)+2)^2$, where the field $E$ is estimated from the relation $cE^2/4\pi = P/S$, with $P$ and $S$ being the power of laser and the area of the light spot. With account of these relations, using Eq. (15) and Newton's second law we obtain (we also bear in mind that the inertia moment of a spherical particle is $8\pi\rho R^5/15$)

$$\frac{d\Omega}{dt} = \frac{15}{8} \frac{R}{z_0} \frac{P}{\rho c S z_0^2} \frac{(\varepsilon(0)-1)^2}{(\varepsilon(0)+2)^2} f(\Omega) \qquad (17)$$

Assuming $R = 1\,nm$, $z_0 = 500\,nm$, $\rho = 2.3\,g/cm^3$, $P = 0.1\,mW = 10^3\,erg/s$ and $S = 0.5\,cm^2$ we obtain from (17) $d\Omega/dt \cong 10^{-2}/f(\Omega)\,(s^{-2})$. Since, typically $f(\Omega) \sim 1$, we thus obtain the characteristic time of the dipole alignment (braking) of order $10^2\,s$, which is feasible to measure. In this case, as it follows from the above consideration, one can expect an increase in the modulus of dipole force (see Eq.(11) and Fig. 2) by 3 to 4 times as compared to the static case $\Omega = 0$, while its sign can be either positive or negative. So, the surface attracts or pushes off the nanoparticle.

## Conclusions

For the first time, we have examined the impact of rotation on the interaction of a small dipole particle with the surface of dielectric. We have calculated the energy of interaction and the torque acting on the particle. As it follows from the obtained expressions, rotation of the particle does not explicitly result in the time dependence of potential energy and $Z-$component of torque in the case where a rotation axis coincides with the surface normal. The torque involves a frictional component, that decelerates general rotation around the $Z$ axis, and an orientational component which tends to orient the dipole axis in the direction normal to surface. The alignment time is always longer than the braking time, so the time interval, when the rotation can be thought of as the stationary process, (provided that the dipole axis is inclined to the surface normal), is determined by the time of dipole orientation. If the dipole axis is directed exactly normal to the surface, the torque is zero and has no impact on rotation and orientation of the dipole, while the force of interaction with the surface coincides with the static (attractive) force. In the case where the dipole axis is inclined at an angle of $\pi/2$ relative to the surface normal, rotation of the particle has a noticeable effect on the interaction force, which increases in modulus (by several times) and changes the sign provided that the frequency of rotation is tuned to absorption resonance of the surface. We have estimated the corresponding effect in the case of a $SiO_2$ nanoparticle above a silica surface. We have shown that the repulsion effect is feasible to measure at typical parameters of the external laser field.

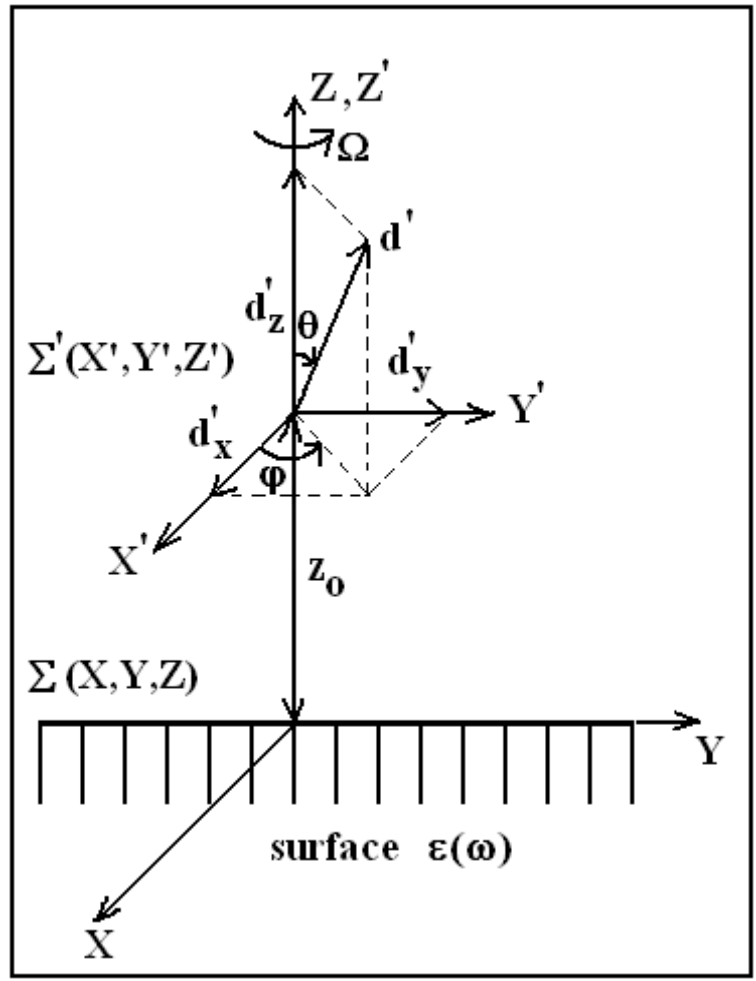

Fig. 1. Configuration of the system and coordinate systems used.

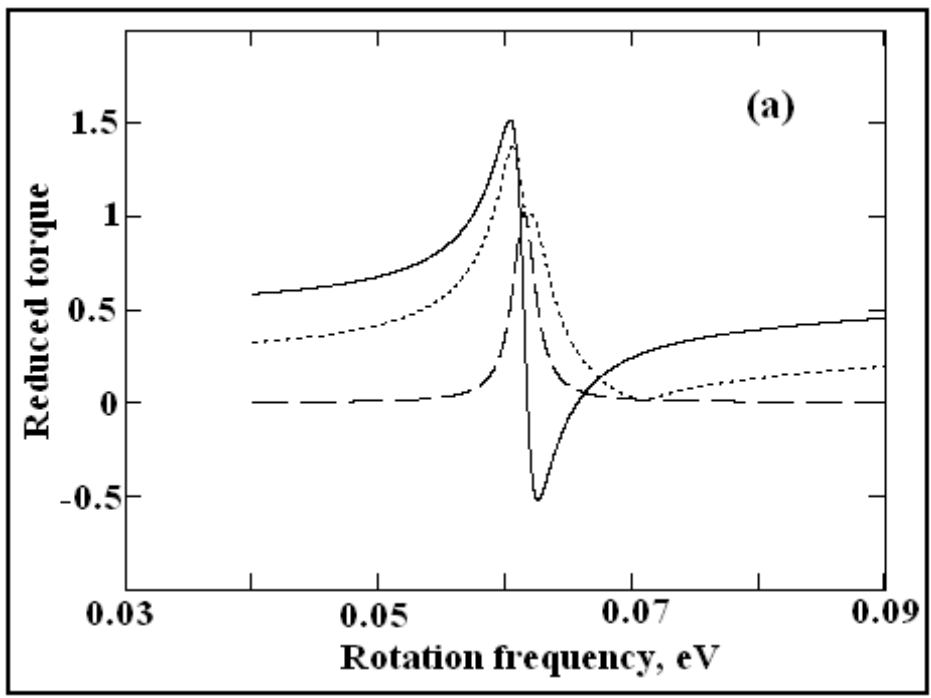

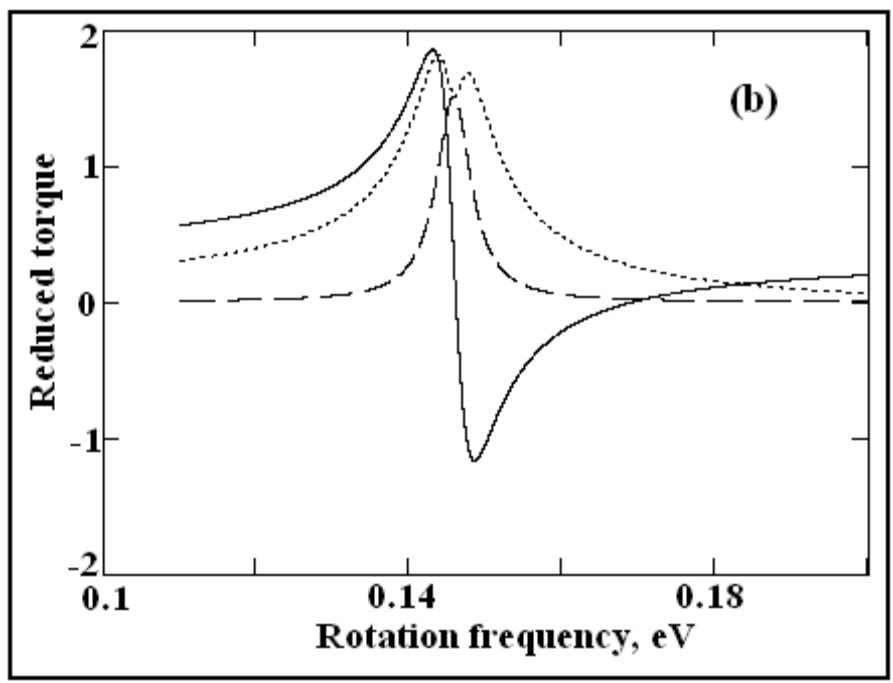

Fig. 2 Reduced torque vs. rotation frequency. Solid line -- $\Delta'(\Omega)$, dashed line $\Delta''(\Omega)$, dotted line— $f(\Omega)$.